

\input harvmac


\overfullrule=0pt


\def\I{{\scriptscriptstyle I}}

\def\L{{\scriptscriptstyle L}}

\def\T{{\scriptscriptstyle T}}


\def\CA{{\cal A}}

\def\CL{{\cal L}}


\def\b{\beta}

\def\e{\epsilon}
\def\g{\gamma}
\def\l{\lambda}

\def\o{\sigma}
\def\th{\theta}

\def\u{\mu}
\def\v{\nu}


\def\aS{\alpha_s}
\def\bar#1{\overline{#1}}

\def\chisq{\chi^2}
\def\Csix{C^{(6)}}
\def\Ceight{C^{(8)}}
\def\costhstar{\cos\th^*}
\def\dash{{\> \over \>}} 		
\def\EFT{{\scriptscriptstyle EFT}}
\def\fourquark{{4 \dash {\rm quark}}}
\def\GeV{{\>\, \rm GeV}}
\def\gfive{\gamma^5}

\def\gs{g_s}
\def\Leff{\CL_{\rm eff}}
\def\LQCD{\CL_{\scriptscriptstyle QCD}}

\def\mperp{{m_\perp}}
\def\mq{{m_q}}
\def\mt{{m_t}}
\def\mtsq{{m_t^2}}
\def\nf{n_f}
\def\obs{{\rm obs}}
\def\Osix{O^{(6)}}
\def\Oeight{O^{(8)}}

\def\pb{\>{\rm pb}}
\def\Pbar{\overline{P}}
\def\pperp{{p_\perp}}
\def\qbar{\overline{q}}
\def\QCD{{\scriptscriptstyle QCD}}
\def\Rang{R_{\rm ang}}
\def\Rpperp{R_{\pperp}}
\def\Rrms{R_{\rm rms}}
\def\shat{\hat{s}}

\def\space{\>\>}
\def\tbar{\overline{t}}
\def\TeV{{\>\, \rm TeV}}
\def\that{\hat{t}}
\def\uhat{\hat{u}}


\def\half{{1 \over 2}}

\def\third{{1 \over 3}}


\newdimen\pmboffset
\pmboffset 0.022em
\def\oldpmb#1{\setbox0=\hbox{#1}%
 \copy0\kern-\wd0
 \kern\pmboffset\raise 1.732\pmboffset\copy0\kern-\wd0
 \kern\pmboffset\box0}


\nref\ssusy{P. Fayet, Nucl. Phys. {\bf B90} (1975) 104 and Phys. Lett.
  {\bf B69} (1977) 489; G.R. Farrar and P. Fayet, Phys. Lett. {\bf B76}
  (1978) 575; E. Witten, Nucl. Phys. {\bf B188} (1981) 513; S.
  Dimopoulos and H. Georgi, Nucl. Phys. {\bf B193} (1981) 150; N. Sakai,
  Z. Phys. {\bf C11} (1981) 153; L. Iba\~nez and G. Ross, Phys. Lett. {\bf
  B105} (1981) 439; R.K. Kaul, Phys. Lett. {\bf B109} (1982) 19; M.
  Dine, W. Fischler and M. Srednicki, Nucl. Phys. {\bf B189} (1981) 575;
  S. Dimopoulos and S. Raby, Nucl. Phys. {\bf B192} (1981) 353.}
\nref\techni{S. Weinberg, Phys. Rev. {\bf D19}, (1979) 1277; L. Susskind,
  Phys. Rev. {\bf D20} (1979) 2619; E. Farhi and L. Susskind, Phys. Rep.
  {\bf 74} (1981) 277.}
\nref\Marciano{W. Marciano, Phys. Rev. {\bf D21} (1980) 2425.}
\nref\compness{See e.g. the review by M. Peskin in {\sl Proceedings of
  the International Symposium on Lepton and Photon Interactions at High
  Energies, Bonn, 1981} edited by W. Pfeil (Physikalisches Institut,
  Universit\"at Bonn, 1981), p. 880}
\nref\ELP{E. Eichten, K. Lane and M. Peskin, Phys. Rev. Lett. {\bf 50}
  (1983) 811.}
\nref\Buchmuller{W. Buchm\"uller and D. Wyler, Nuc. Phys. {\bf B268} (1986)
  621.}
\nref\CDF{CDF Collaboration, F. Abe {\it et al.}, Phys. Rev. Lett. {\bf 68}
  (1992) 1104.}
\nref\SimmonsI{E.H. Simmons, Phys. Lett. {\bf B226} (1989) 132.}
\nref\Politzer{H.D. Politzer, Nucl. Phys. {\bf B172} (1980) 349.}
\nref\SimmonsII{E.H. Simmons, Phys. Lett. {\bf B246} (1990) 471.}
\nref\Morozov{A.Y. Morozov, Sov. J. Nucl. Phys. {\bf 40} (1984) 505.}
\nref\Dreiner{H. Dreiner, A. Duff and D. Zeppenfeld, Phys. Lett. {\bf B282}
  (1992) 441.}
\nref\Duff{A. Duff and D. Zeppenfeld, Z. Phys. {\bf C53} (1992) 529.}
\nref\Dixon{L. Dixon and Y. Shadmi, SLAC-PUB-6416 (1993).}
\nref\Narison{S. Narison and R. Tarrach, Phys. Lett. {\bf B125} (1983) 217.}
\nref\topmass{CDF Collaboration, F. Abe {\it et al.}, FERMILAB-PUB-94/097-E
  (1994); FERMILAB-PUB-94/116-E (1994).}
\nref\Pietrzyk{B. Pietrzyk, Talk presented at the XXIXth Rencontres de
  Moriond, M\'eribel, March 1994, LAPP-EXP-94.07}
\nref\ChoSimmons{P. Cho and E. Simmons, Phys. Lett. {\bf B323} (1994) 401.}
\nref\Combridge{M.Gluck, J.F. Owens and E. Reya, Phys. Rev. {\bf D17}
(1978) 2324\semi B.L. Combridge, Nucl. Phys. {\bf B151} (1979) 429.}
\nref\Ellis{R.K. Ellis, in {\it Proceedings of the 17th SLAC Summer
  Institute}, Stanford, Calif., 1989, edited by E.C. Brennan, SLAC Rept. 361,
  45.}
\nref\Harriman{P. Harriman, A. Martin, R. Roberts and J. Stirling, Phys.
 Rev. {\bf D42} (1990) 798.}
\nref\Laenen{E. Laenen, J. Smith and W.L. Van Neerven, Nucl. Phys. {\bf B369}
 (1992) 543; {\it ibid}, FERMILAB-Pub-93/270-T.}
\nref\PDB{Review of Particle Properties, Particle Data Group, Phys. Rev. {\bf
 D45} (1992) III.38.}


\nfig\Loopgraphs{Diagrams with a heavy colored fermion or boson running around
the loop which match onto gluonic operators in the effective strong interaction
Lagrangian.}
\nfig\ggttbargraphs{(a) Lowest order QCD graphs which contribute to
$gg \to t\tbar$ scattering.  (b) $O(1/\Lambda^2)$ $gg \to t\tbar$ graphs with
single insertions of the chromomagnetic moment operator $\Osix_0$.
(c) $O(1/\Lambda^2)$ $gg \to t\tbar$ graph with single insertion of the triple
gluon field strength operator $\Osix_1$. (d) $O(1/\Lambda^4)$ $gg \to t\tbar$
graphs with two insertions of either $\Osix_0$ or $\Osix_0$ and $\Osix_1$.}
\nfig\qqbarttbargraphs{Lowest order QCD, chromomagnetic moment and four-quark
operator graphs which contribute to $q\qbar \to t\tbar$ scattering.}
\nfig\LHCpTplot{Transverse momentum differential cross section
$d\sigma(PP \to t\tbar)/d\pperp$ calculated for an LHC center-of-mass energy
$\sqrt{s} = 14 \TeV$.  The solid curve illustrates the pure QCD cross section.
The dot-dashed, dashed and dotted curves represent the additional
nonrenormalizable operator contributions obtained after respectively setting
$\Csix_0(\Lambda)$, $\Csix_1(\Lambda)$ and $\Csix_2(\Lambda)$ equal to 0.5
with $\Lambda=2\TeV$.}
\nfig\FNALpTplot{Transverse momentum differential cross section
$d\sigma(P{\bar P} \to t\tbar)/d\pperp$ calculated for a Tevatron
center-of-mass energy $\sqrt{s} = 1.8 \TeV$.  The curves are labeled the
same as those in \LHCpTplot.  The dashed $\Osix_1$ curve must be subtracted
from rather than added to the solid QCD curve to obtain the EFT cross section
that corresponds to $\Csix_1(\Lambda)=0.5$.}
\nfig\Rplot{Ratio $\Rpperp$ of the EFT and QCD $t\tbar$ transverse momentum
distributions integrated over $500 \GeV \le \pperp \le 1000 \GeV$
plotted as a function of gluonic operator coefficients $\Csix_1(\Lambda)$ and
$\Csix_2(\Lambda)$ with $\Lambda=2\TeV$ and $\sqrt{s}=14 \TeV$.  The values
for $\Rpperp$ are displayed alongside the contours.}
\nfig\chisqplot{$\chisq/N$ for $N=20$ transverse momentum bins plotted as
a function of gluonic operator coefficients $\Csix_1(\Lambda)$ and
$\Csix_2(\Lambda)$ with $\Lambda=2\TeV$ and $\sqrt{s}=14\TeV$.  The innermost
crescent contour corresponds to $\chisq/N=1$.  The surrounding contours
represent $\chisq/N = 1 + 2n\sigma$ where $\sigma=\sqrt{2/N}=0.316$ and
$n=1,2,3,4$.}
\nfig\angleplot{Angular differential cross section $d\sigma(PP \to
t\bar t)/d\cos\theta^*$ calculated for $\sqrt{s} = 14$ TeV.  The solid
curve illustrates the pure QCD cross section.  The dotted curve corresponds
to QCD plus the dimension-6 operator $\Osix_1$ with $\Csix_1(\Lambda)=0.5$ and
$\Lambda=2\TeV$.  The dashed and dot-dashed curves represent QCD plus the
dimension-8 operator $\Oeight_3$ with $\Ceight_3(\Lambda)$ set equal to 0.5
and -0.5 respectively.}


\def\CITTitle#1#2#3{\nopagenumbers\abstractfont
\hsize=\hstitle\rightline{#1}
\vskip 0.3in\centerline{\titlefont #2} \centerline{\titlefont #3}
\abstractfont\vskip .2in\pageno=0}

\CITTitle{{\baselineskip=12pt plus 1pt minus 1pt
  \vbox{\hbox{CALT-68-1941}\hbox{DOE RESEARCH AND}\hbox{DEVELOPMENT REPORT}
  \hbox{BUHEP-94-18}}}}
  {Searching for $G^3$ in $t\tbar$ Production}{}
\centerline{Peter Cho\footnote{$^*$}{Work supported in part by an SSC
 Fellowship and by the U.S. Dept. of Energy under DOE Grant no.
 DE-FG03-92-ER40701.}}
\centerline{Lauritsen Laboratory}
\centerline{California Institute of Technology}
\centerline{Pasadena, CA  91125}
\medskip\centerline{and}\medskip
\centerline{Elizabeth H. Simmons\footnote{$^{**}$}{Work supported in
part by an American Fellowship from the American Association of
University Women.}}
\centerline{Department of Physics}
\centerline{Boston University}
\centerline{590 Commonwealth Avenue}
\centerline{Boston, MA\ 02215}

\vskip .15in
\centerline{\bf Abstract}
\bigskip

	The triple gluon field strength operator $G^3$ represents the only
genuinely gluonic CP conserving term which can appear at dimension-6 within an
effective strong interaction Lagrangian.  Previous studies of this operator
have revealed that its effect on gluon scattering is surprisingly difficult to
detect.  In this article, we analyze the impact of $G^3$ upon top quark pair
production.  We find that it will generate observable cross section deviations
from QCD at the LHC for even relatively small values of its coefficient.
Furthermore, $G^3$ affects the transverse momentum distribution of the
produced top quarks more strongly at high energies than dimension-6
four-quark and chromomagnetic moment terms in the effective Lagrangian.
Top-antitop production at the LHC will therefore provide a sensitive and clean
probe for the elusive triple gluon field strength operator.

\Date{8/94}

\newsec{Introduction}

	The quest to uncover signs of new physics beyond the Standard
Model is being pursued across a broad front.  One important area in
which the search has been underway for many years is the strong
interaction sector.  Quantum chromodynamics has so far passed every
experimental test and explains all strong interaction phenomenology at
currently accessible energies.  However, new color interactions beyond
those of conventional QCD could emerge at higher energy scales from a
number of different sources.  For example, a wide range of theories
beyond the Standard Model contain novel particles which couple to
quarks and gluons. Such particles could be squarks and gluinos in
supersymmetric theories \ssusy, colored technihadrons in non-minimal
technicolor models \techni, or fermions that transform according
to representations of color $SU(3)$ other than the fundamental triplet in
certain electroweak symmetry breaking scenarios \Marciano.  New strong sector
physics might alternatively originate from quark or gluon substructure
\compness.  Preon exchange between composite states would generally induce
nonstandard quark or gluon couplings.  In short, the variety of novel color
interactions which have been proposed in the literature provides many
possible sources of interesting physics.

	A complete description of any new strong interaction phenomena
requires a fundamental theory beyond the Standard Model.  However
at energies below its characteristic scale $\Lambda$, the new physics
may be studied within an effective field theory framework.  Its low energy
effects can be reproduced by supplementing the renormalizable terms in the
QCD Lagrangian with higher dimension operators.  The resulting effective
Lagrangian
\eqn\effLagrange{\Leff = \LQCD + {1 \over \Lambda^2} \sum_i \Csix_i(\mu)
\Osix_i(\mu) + {1\over \Lambda^4} \sum_i \Ceight_i(\mu) \Oeight_i(\mu) +
O\bigl({1 \over \Lambda^6} \bigr)}
generally contains all terms consistent with $SU(3) \times SU(2) \times U(1)$
gauge invariance and any imposed discrete or global symmetries.
The operators $O_i^{(n)}$ and their dimensionless coefficients $C_i^{(n)}$
are grouped together in eqn.~\effLagrange\ according to their mass
dimension $n$ and multiplied by a factor of $1/\Lambda^{n-4}$ so
that their overall combined dimension equals 4.  As a result, the most
important nonrenormalizable terms in $\Leff$ are those of dimension-6
since they are least suppressed by inverse powers of the high energy
scale $\Lambda$.

	In the quark sector, one can form a large number of dimension-6
operators with different chiral and color structures \refs{\ELP,\Buchmuller}.
The qualitative impact of these terms upon quark scattering has conventionally
been assessed by including just one representative four-quark operator such as
\eqn\fourquarkop{\Osix_\fourquark =
{1 \over 2} (\bar{q}_\L \g^\u q_\L) (\bar{q}_\L \g_\u q_\L)}
into the effective Lagrangian and setting the magnitude of its coefficient
equal to $4 \pi$.  The scale $\Lambda$ associated with the four-quark operator
has then been constrained by comparing its effect upon the theoretical
prediction for the inclusive jet cross section with experimental measurement.
A recent such comparison yields the bound $\Lambda > 1.4 \TeV$ \CDF.

	The number of nonrenormalizable terms which arise at dimension-6
in the gluon sector is much more limited.  One can build only two gauge
invariant operators out of covariant derivatives
$D^\u = \partial^\u - i \gs G^\u_a T_a$ and gluon field strengths
$G^{\u\v}_a = \partial^\u G^\v_a - \partial^\v G^\u_a +
\gs f_{abc}G^\u_b G^\v_c$ that preserve $C$, $P$ and $T$ \SimmonsI:
\eqna\dimsixops
$$ \eqalignno{
\Osix_1 &= \gs f_{abc} G^\u_{a\v} G^\v_{b\l} G^\l_{c\u} &
\dimsixops a \cr
\Osix_2 &= {1 \over 2} D^\u G^a_{\u\v} D_\l G^{\l\v}_a. &
\dimsixops b \cr} $$
Other candidates such as $\gs d_{abc} G^\u_{a\v} G^\v_{b\l} G^\l_{c\u}$ or
$\half D^\l G^a_{\u\v} D_\l G^{\u\v}_a $ either vanish or reduce to
combinations of $\Osix_1$ and $\Osix_2$.   The triple gluon field strength
term in \dimsixops{a}, which we shall name $G^3$ for short, represents a
true gluonic operator.  It contributes for instance to the distinctive
three-point vertex in the nonabelian color theory.  On the other hand, the
double gluon field strength operator in \dimsixops{b}, which we will call
$(DG)^2$, is not really gluonic.  The classical equation of motion
\eqn\EOM{D_\u G^{\u\v}_a = -\gs \sum_{\rm flavors} \qbar \g^\v T_a q}
relates its S-matrix elements to those of a color octet four-quark operator
\Politzer:
\eqn\Smatrixreln{\Osix_2
\> {\buildrel {\scriptscriptstyle EOM} \over \longrightarrow} \>
{\gs^2\over 2} \sum_{\rm flavors}
\bigl( \qbar \g_\u T_a q  \bigr) \bigl( \qbar \g^\u T_a q \bigr).}
The double field strength operator thus affects parton processes involving
external quarks rather than external gluons.  While the origin of $\Osix_2$
could significantly differ from those of  other four-quark operators, its
impact upon parton scattering is not very different from theirs.  So
at dimension-6, only one genuinely gluonic term which conserves $CP$ may
appear within the effective Lagrangian.

	The list of $CP$ even gluon operators grows at dimension-8 but remains
manageable in size.  It is useful to classify them according to the number
of field strengths that they contain \SimmonsII.  Only one independent
operator can be built out of two field strengths and four covariant
derivatives:
\eqna\dimeightops
$$ \eqalignno{\Oeight_1 &= {1 \over 2} D^\u G^a_{\u\v} D^2 D_\l G^{\l\v}_a.
& \dimeightops a}$$
Any other candidate in this category such as
$\half D^2 G^a_{\u\v} D^2 G_a^{\u\v}$ may be reduced via the Jacobi identity
$D_\l G^a_{\u\v} + D_\u G^a_{\v\l} + D_\v G^a_{\l\u} = 0$ to $\Oeight_1$ plus
operators with more than two field strengths.  There are two possibilities for
dimension-8 operators with three gluon field strengths and two covariant
derivatives:
\foot{Operator $\Oeight_3$ does not reduce to $\Oeight_2$ plus a four
field strength operator as previously reported in ref.~\SimmonsII.}
$$ \eqalignno{\eqalign{
\Oeight_2 &= \gs f_{abc} G^\u_{a\v} D_\l G^{\l\v}_b D^\o G^c_{\o\u} \cr
\Oeight_3 &= \gs f_{abc} G^\u_{a\v} G^\v_{b\l} D^2 G^\l_{c\u}. \cr}
&& \dimeightops b} $$
Finally, six independent operators containing four field strengths
can be formed \Morozov:
$$ \eqalignno{\eqalign{
\Oeight_4 &= {\gs^2 \over 2} G^a_{\u\v} G^{\u\v}_a G^b_{\l\o}
	      G^{\l\o}_b \cr
\Oeight_5 &= {\gs^2 \over 2} G^a_{\u\v} \tilde{G}^{\u\v}_a G^b_{\l\o}
             \tilde{G}^{\l\o}_b \cr
\Oeight_6 &= {\gs^2 \over 2} G^a_{\u\v} G^{\u\v}_b G^a_{\l\o}
	     G^{\l\o}_b \cr
\Oeight_7 &= {\gs^2 \over 2} G^a_{\u\v} \tilde{G}^{\u\v}_b
       	     G^a_{\l\o} \tilde G^{\l\o}_b \cr
\Oeight_8 &= {\gs^2 \over 2} d_{abe} d_{cde} G^a_{\u\v} G^{\u\v}_b
	     G^c_{\l\o} G^{\l\o}_d \cr
\Oeight_9 &= {\gs^2 \over 2} d_{abe} d_{cde} G^a_{\u\v} \tilde{G}^{\u\v}_b
	     G^c_{\l\o} \tilde{G}^{\l\o}_d. \cr}
&& \dimeightops c} $$
The equation of motion in \EOM\ relates the S-matrix elements of $\Oeight_1$
and $\Oeight_2$ to those of operators with four quark fields.  So among
all the dimension-8 operators, only $\Oeight_3$ - $\Oeight_9$ are genuinely
gluonic and affect scattering processes with external gluons.

	Present day experiments at the Fermilab Tevatron are more
sensitive to the presence of nonrenormalizable quark operators in the
strong interaction effective Lagrangian than to their gluonic counterparts.
But at higher energy machines such as the LHC, the gluon content of
colliding hadrons at small parton momenta fractions dominates over that
of valence and sea quarks.  Future collider experiments will therefore probe
possible deviations from the Standard Model much more sensitively
in the gluon rather than quark sectors.   In this article, we will investigate
the prospects for detecting such deviations.  In particular, we will focus
upon searching for the triple gluon field strength operator $G^3$.

	The most obvious channel in which to look for $G^3$
might seem to be $gg \to gg$ scattering.  However, the helicity structure
of the amplitude for this process which involves the gluon operator is
orthogonal to that of pure QCD.   They consequently do not interfere at
$O(1/\Lambda^2)$ \SimmonsI.  At $O(1/\Lambda^4)$, the dimension-8 operators
$\Oeight_3$ - $\Oeight_9$ enter into the $gg \to gg$ amplitude and contaminate
any signal from $\Osix_1$ \SimmonsII.  So looking for $G^3$ in lowest order
gluon scattering is difficult.  As we shall later see, the triple field
strength operator does not mix at one-loop order with any four-quark
operators.  Its presence in the effective Lagrangian thus cannot be indirectly
inferred from pure quark processes either.

	These obstacles to observing $G^3$ have motivated consideration of
other possible detection methods.   For instance, it may be probed
in $Z \to 4 \space{\rm jets}$ decay at LEP \refs{\Dreiner,\Duff}.  This
option is limited, though, by LEP's relatively low energy.
Alternatively, one can look for the effect of $G^3$ upon $gg \to ggg$
scattering in three-jet events \Dixon.  The task of identifying
its signature in this next-to-leading order channel represents a
formidable experimental challenge.  The triple gluon field strength operator
has thus remained stubbornly difficult to detect.

	In this paper, we take a new approach to searching for the $G^3$
operator which possesses significant advantages over those previously
explored.  Specifically, we investigate the effect of the triple field
strength term upon top quark pair production.  As we shall demonstrate, the
impact of $G^3$ at LHC energies is sizable for even relatively small values of
its effective Lagrangian coefficient.  It dominates in this mode over
four-quark competitors with comparable coefficients.  Moreover, contamination
from gluon terms of dimension-8 and higher is small.  So $t\tbar$ production
provides a sensitive and clean probe for the elusive $G^3$ operator.

	Our article is organized as follows.  In section 2, we discuss the
renormalization group running of all the dimension-6 operator coefficients that
influence top quark pair production.  We then compute in section 3 the
$t\tbar$ differential transverse momentum cross sections at the LHC and
Tevatron for representative values of their operator coefficients and compare
with corresponding QCD results.  In section 4, we quantify the difference
between the transverse momentum cross sections calculated with and without the
gluonic operators in the effective Lagrangian as a function of their
coefficients.  In section 5, we investigate the operators' impact upon
$t\tbar$ angular distributions.  Finally, we summarize our findings and
present our conclusions in section 6.

\newsec{Dimension-6 operator coefficients}

	Many forms of physics beyond the Standard Model could give
rise to nonrenormalizable gluonic operators within the effective
strong interaction Lagrangian.  In the absence of a fundamental
theory, the coefficients of such operators are {\it a priori} unknown.
Even when a particular high energy model is specified, their values
may prove intractably difficult to calculate.  So in this work, we
will simply study the feasibility of detecting the lowest dimension
operators as a function of their coefficient values.

	Since the dimensionless coefficients and dimensionful scale
appear together within the effective Lagrangian \effLagrange, they
cannot be distinguished at tree level.  We are therefore free to fix
the value for the high energy scale.  Motivated by the expectation
that a new fundamental layer of physics awaits discovery in the
TeV regime, we will take it to be $\Lambda = 2 \TeV$.  This relatively
large value for the effective theory cutoff ensures that our low
energy analysis will be valid over almost the entire practical energy range
of present and anticipated hadron colliders.

	Although we will generally treat the coefficients of the gluon
operators as free parameters, we note that there is an important special case
in which they can readily be computed.  This occurs when the operators are
generated through diagrams involving heavy colored particle loops like those
illustrated in \Loopgraphs.  At energies below the particle's mass, the
behavior of such graphs may be reexpressed in terms of local but
nonrenormalizable gluonic operators.  The values for their coefficients are
then determined by performing a matching computation.  For example, if the
particles running around the loop in \Loopgraphs\ are fermions of mass
$\Lambda$ and belong to representation $R$ of color $SU(3)$, the sum of the
diagrams contains terms which match onto the dimension-6 operators in
eqn.~\dimsixops{}\ with coefficients
\eqn\fermionmatching{\eqalign{
\Csix_1(\Lambda) &= -{K(R) \over 360} {\aS(\Lambda) \over \pi} \cr
\Csix_2(\Lambda) &= -{K(R) \over 15} {\aS(\Lambda) \over \pi} \cr}}
where $K(R)=\sum_{a=1}^8 \Tr[T_a(R) T_a(R)]/8$ denotes the index of $R$.
The corresponding result for colored scalars is given by
\eqn\bosonmatching{\eqalign{
\Csix_1(\Lambda) &= {K(R) \over 720} {\aS(\Lambda) \over \pi} \cr
\Csix_2(\Lambda) &= -{K(R) \over 120} {\aS(\Lambda) \over \pi}. \cr}}
Unfortunately, the numerical values of these coefficients are too small in any
reasonable theory to produce detectable deviations from the Standard
Model.  We will therefore have to be content with
constraining mechanisms other than perturbative radiative corrections which
can generate gluonic operators with larger coefficients.

	Whatever values $\Csix_1$ and $\Csix_2$ may assume
at the scale $\Lambda$, they must be evolved down to the energies at which
$\Osix_1$ and $\Osix_2$ are probed using the renormalization group.  We
consequently need to enlarge our operator basis so that it closes under
renormalization.  It is sensible to apply the equations of motion to
reduce the operator set as much as possible.  We then find that $\Osix_1$ and
$\Osix_2$ do not mix with each other under the action of QCD at one-loop
order.  Instead, $\Osix_1$ runs into itself and the chromomagnetic moment
operator
\eqn\magmomop{\Osix_0 = \sum_{\rm flavors} \gs \mq
  \qbar \o^{\u\v} T^a q G^a_{\u\v}.}
Their evolution is governed by the $2 \times 2$ anomalous dimension
matrix
\eqn\anomdimmatrixI{\g_\I = \bordermatrix{& \Osix_0 & \Osix_1 \cr
\Osix_0 & 14/3 & 0 \cr
\Osix_1 & 9/2 & 7+{2\nf/3} \cr} {\gs^2 \over 8\pi^2} + O(\gs^4)}
where $\nf$ denotes the number of active quark flavors
\refs{\Morozov,\Narison}.  In the $\Osix_2$ sector, the equations of motion
convert the double gluon field strength operator into the color octet
four-quark term in \Smatrixreln.  $\Osix_2$ then mixes at one-loop order
with its four-quark counterparts
\eqn\quarkoplist{\eqalign{
\Osix_3 &= {\gs^2\over 2} \sum_{\rm flavors}
   \bigl( \qbar \g_\u \gfive T_a  q \bigr)
   \bigl( \qbar \g^\u \gfive T_a q \bigr) \cr
\Osix_4 &= {\gs^2\over 2} \sum_{\rm flavors}
   \bigl( \qbar \g_\u q \bigr) \bigl( \qbar \g^\u q \bigr) \cr
\Osix_5 &= {\gs^2\over 2} \sum_{\rm flavors}
   \bigl( \qbar \g_\u \gfive q \bigr) \bigl( \qbar \g^\u \gfive q \bigr) \cr}}
through the $4 \times 4$ matrix
\eqn\anomdimmatrixII{\g_{\I\I} = \bordermatrix{& \Osix_2 & \Osix_3 & \Osix_4
& \Osix_5 \cr
\Osix_2 & 311/36-2\nf/3 & 5/4 & 0 & 2/3 \cr
\Osix_3 & 41/36 & 35/4-2\nf/3 & 2/3 & 0 \cr
\Osix_4 & 4/3 & 6 & 11-{2\nf/3} & 0 \cr
\Osix_5 & 22/3 & 0 & 0 & 11-{2\nf/3} \cr} {\gs^2 \over 8\pi^2}+O(\gs^4).}

	The coefficients of operators $\Osix_0$ - $\Osix_5$ satisfy the
integrated renormalization group equation
\eqn\RGEsoln{C_i(\mu) = \sum_j \Bigl[ \exp \int^{g_s(\mu)}_{g_s(\Lambda)}
dg_s {\gamma^\T (g_s) \over \beta(g_s)} \Bigr]_{ij} C_j(\Lambda)}
where $\b(g_s)$ denotes the QCD beta function.
\foot{We take the constant of integration which enters into the integrated
QCD beta function to be $\aS(M_z) = 0.125$ \Pietrzyk.  The corresponding
value for the strong interaction fine structure constant at $\Lambda=2\TeV$
is $\aS(\Lambda)=0.086$ which implies $g_s(\Lambda)=1.04$.  Since this
coupling is very close to unity, the value of $\Csix_1(\Lambda)$ is
essentially independent of the number of powers of $g_s$ that are included
into the $G^3$ operator's definition.}
The leading log running of these coefficients can be substantial.  For
example, if they assume the values
\eqn\coeffshi{
(\Csix_0,\Csix_1,\Csix_2,\Csix_3,\Csix_4,\Csix_5)(\Lambda) = (1,1,1,0,0,0)}
at the canonical $\Lambda=2 \TeV$ scale, the coefficients run down to
\eqn\coeffslo{\eqalign{
& (\Csix_0,\Csix_1,\Csix_2,\Csix_3,\Csix_4,\Csix_5)(2\mt) = \cr
&\qquad\qquad (0.7858, 0.7458, 0.8856, -0.0294, 0.0003, -0.0152)}}
at the top-antitop threshold.
\foot{We adopt the recently reported CDF central value $\mt = 174 \GeV$ for
the mass of the top quark \topmass.}
Renormalization group evolution therefore generally suppresses the operator
coefficients.

\newsec{Top quark pair production}

	The sensitivity of a hadron collider experiment to  nonrenormalizable
terms in the effective strong interaction Lagrangian is a function of the
accelerator's center-of-mass energy $\sqrt{s}$ and luminosity $\CL$.  The
Fermilab Tevatron currently operates at $\sqrt{s}=1.8 \TeV$ and has collected
approximately 20 ${\rm pb}^{-1}$ of data.  These values for $\sqrt{s}$ and
$\int \CL dt$ are too low to allow for any significant limit to be placed upon
the coefficient of the triple gluon field strength operator $G^3$.  However,
the prospects for probing $G^3$ will substantially improve with the advent of
the LHC.  This machine is projected to run at $\sqrt{s} = 14 \TeV$ and collect
approximately 30 ${\rm fb}^{-1}$ of data per year.  At such high energies, the
gluon content of the colliding hadrons at small parton momenta fractions
dominates over that of all other partons.  Sensitivity to the triple field
strength operator will consequently be greatly enhanced.

	As mentioned in the introduction, looking for $G^3$ in $gg \to gg$
scattering is problematic.  The only other $2 \to 2$ parton process that
maximally benefits from the large numbers of gluons in a high $\sqrt{s}$
initial state is $gg \to q\qbar$.  The $O(1/\Lambda^2)$ interference between
the QCD and $G^3$ amplitudes for this mode is proportional to the squared
quark mass $\mq^2$.  It is therefore greatest when the final state quark
flavor is top.  The top quark is also the easiest to tag since its leptonic
decay channel can produce a high energy, isolated lepton in conjunction with a
bottom quark.  This distinctive signature cuts down on genuine backgrounds
as well as false identifications.  If in the future $b\bar{b}$ and even
$c\bar{c}$ final states can be positively identified, then the signal for
$G^3$ will only be enhanced and all our results can be easily applied to
their study.  But for now, we shall examine the impact of the
nonrenormalizable terms in $\Leff$ upon just top quark pair production.  As
we shall see, this mode provides a sensitive and clean probe for the
$G^3$ operator.

	Top quark pair production proceeds at tree level through the parton
reactions $gg \to t\tbar$ and $q \qbar \to t\tbar$.  We first consider the
gluon channel.  The lowest order QCD graphs that mediate $gg \to
t\tbar$ scattering are illustrated in \ggttbargraphs a.  At $O(1/\Lambda^2)$,
the chromomagnetic moment and triple gluon field strength operators $\Osix_0$
and $\Osix_1$ contribute through the diagrams shown in \ggttbargraphs b
and \ggttbargraphs c.  They also enter into the $gg \to t\tbar$ amplitude at
$O(1/\Lambda^4)$ via the double operator insertion graphs displayed in
\ggttbargraphs d.  The only other possible dimension-6 operator tree diagrams
contain insertions of the double field strength operator $\Osix_2$.  But the
sum of such $(DG)^2$ graphs vanishes as guaranteed by the S-matrix relation
in \Smatrixreln.
\vfill\eject

	After adding together all the diagrams in \ggttbargraphs\ and squaring
the total $gg \to t\tbar$ amplitude, we find
\foot{The results displayed here in eqns.~(3.1) and (3.3) correct errors
in eqns.~(14a) and (14b) of ref.~\ChoSimmons\ and eqns.~(4.4) and (4.5) of
ref.~\SimmonsII . The pure QCD terms in these formulae agree with the results
of ref.~\Combridge.}
\eqn\gluesqrdamp{\eqalign{\mathop{{\bar\sum}'} | \CA(gg \to t\tbar) |^2 &=
{3 \over 4} {(\mtsq-\that)(\mtsq-\uhat) \over \shat^2} -{1\over 24}
   {\mtsq(\shat-4\mtsq) \over (\mtsq-\that)(\mtsq-\uhat)} \cr
& + {1\over 6} \Bigl[ {\that\uhat-\mtsq(3\that+\uhat) -\mt^4 \over
   (\mtsq-\that)^2} + {\that\uhat-\mtsq(\that+3\uhat)-\mt^4 \over
   (\mtsq-\uhat)^2} \Bigr] \cr
& -{3 \over 8} \Bigl[ {\that\uhat-2\mtsq\that+\mt^4 \over
   \shat (\mtsq-\that)} + {\that\uhat-2\mtsq\uhat+\mt^4 \over
   \shat(\mtsq-\uhat)} \Bigr] \cr
& + {1 \over \Lambda^2} \Bigr[ \third \Csix_0 {\mtsq
   (4 \shat^2-9 \that\uhat-9\mtsq\shat+9 \mt^4) \over
   (\mtsq-\that) (\mtsq-\uhat)} \cr
&\qquad +  {9 \over 8} \Csix_1 {\mtsq (\that-\uhat)^2 \over
   (\mtsq-\that)(\mtsq-\uhat)} \Bigr] \cr
&+{1 \over \Lambda^4} \Bigl[
   {1 \over 6} {\Csix_0}^2 {\mtsq \bigl( 14 \shat\that\uhat
   + \mtsq(31 \shat^2-36 \that\uhat) - 50 \mt^4 \shat + 36 \mt^6 \bigr) \over
   (\mtsq-\that)(\mtsq-\uhat)} \cr
&\qquad +{9 \over 8} \Csix_0 \Csix_1 {\mtsq \shat^3 \over
   (\mtsq-\that)(\mtsq-\uhat)}
+ {27 \over 4} {\Csix_1}^2 (\mtsq-\that)(\mtsq-\uhat) \Bigr]
+ O\Bigl({1 \over \Lambda^6} \Bigr). \cr}}
The bar appearing over the summation symbol on the LHS of \gluesqrdamp\
implies that the squared matrix element is averaged (summed) over initial
(final) spins and colors, while the prime indicates that $|A(gg \to t\tbar)|^2$
is divided by $\gs^4$.  Notice that all of the nonrenormalizable operator
terms except the last one are proportional to $\mtsq$.  The corresponding
terms in the squared $gg \to q\qbar$ amplitude for quark flavors other than top
are therefore negligible by comparison.  The last term in \gluesqrdamp\ is
significantly enhanced by its $27/4$ prefactor.  Moreover, it increases
quadratically in the partonic Mandelstam invariants $\shat$, $\that$ and
$\uhat$.  In contrast, the other $O(1/\Lambda^4)$ terms only grow linearly,
while the $O(1/\Lambda^2)$ terms approach a constant.  So away from the
$t\tbar$ threshold and over large regions of $\Csix_1$ parameter space,
the $\Osix_1$ operator's squared amplitude is much larger than its
interference with QCD.  As a result, the impact of $G^3$ upon
$t\tbar$ production depends mainly upon the magnitude rather than sign of
its coefficient.

	Since the gluonic term proportional to ${\Csix_1}^2$ in \gluesqrdamp\
is surprisingly large, one may question whether other $O(1/\Lambda^4)$
terms arising from dimension-8 gluon operators could be significant as well.
The answer is generally no.  Recall that among the dimension-8 operators
listed in eqn.~\dimeightops{}, $\Oeight_1$ and $\Oeight_2$ enter at tree
level into processes involving at least four quarks, while vertices from
$\Oeight_4 \dash \> \Oeight_9$ contain at least four gluons.  So only
$\Oeight_3$ can affect $g g \to t\tbar$ scattering at lowest order.  The
interference between its amplitude and that of pure QCD yields
the $O(1/\Lambda^4)$ term
\eqn\Oeightsqrdamp{\mathop{{\bar\sum}'} | \CA(gg \to t\tbar) |^2 = \cdots
-{3 \over 8} {\Ceight_3 \over \Lambda^4} {\mtsq \shat(\that-\uhat)^2
\over(\mtsq-\that)(\mtsq-\uhat)}.}
This dimension-8 term has a much smaller prefactor and increases more slowly
with $\shat$, $\that$ and $\uhat$ than its dimension-6 competitor.  So unless
its coefficient $\Ceight_3$ is more than an order of magnitude larger than
$\Csix_1$, the $\Oeight_3$ operator is not likely to obscure any signal from
$\Osix_1$.

	We now turn to consider the quark process $q\qbar \to t\tbar$.
The chromomagnetic moment, double gluon field strength and four-quark
operators in our basis contribute at lowest order in the strong interaction
coupling to this channel as indicated in \qqbarttbargraphs.  Neglecting all
quark masses except that of the top in the amplitude sum, we find
\eqn\quarksqrdamp{\eqalign{
\mathop{{\bar\sum}'} | \CA(q \qbar \to t\tbar) |^2 &= {4 \over 9 \shat^2}
  \bigl[ \that^2+\uhat^2+4\mtsq\shat-2 \mt^4 \bigr] \cr
& + {8\over 9 \shat \Lambda^2}\bigl[4 \Csix_0 \mtsq \shat +
\Csix_2 (\that^2+\uhat^2+4\mtsq\shat-2\mt^4)
  + \Csix_3 \shat (\that-\uhat) \bigr] \cr
& + { 4 \over 9 \Lambda^4} \Bigl[ 8 {\Csix_0}^2 \mtsq
(\that\uhat+2\mtsq\shat-\mt^4)/\shat \cr
& \quad\qquad + 8 \Csix_0 \Csix_2 \mtsq \shat
 + 8 \Csix_0 \Csix_3 \mtsq(\that-\uhat) \cr
& \quad\qquad + ({\Csix_2}^2+ \half {\Csix_4}^2)
  (\that^2+\uhat^2+4\mtsq\shat-2 \mt^4) \cr
& \quad\qquad + ({\Csix_3}^2+\half {\Csix_5}^2)
  (\that^2+\uhat^2-2 \mt^4 ) \cr
& \quad\qquad + (2\Csix_2 \Csix_3 + \Csix_4 \Csix_5) \shat(\that-\uhat)
\Bigr]. \cr}}
Unlike the gluonic scattering result in \gluesqrdamp, this expression contains
no anomalously large $O(1/\Lambda^4)$ term.  So we expect that the effect
of dimension-8 and higher operators upon $q\qbar \to t\tbar$ scattering is
small.

	The squared amplitudes in \gluesqrdamp\ and \quarksqrdamp\
enter into the partonic differential cross section
\eqn\ggttbarXsect{
{d \sigma(ab \to t\tbar) \over d\that} = {\pi \aS^2 \over \shat^2}
\mathop{{\bar\sum}'} | \CA(ab \to t \tbar) |^2}
which appears in the full hadronic differential cross section for
top-antitop production \Ellis:
\eqn\ABttbarXsect{
{d^3\sigma \over dy_3 dy_4 d\pperp} \bigl(AB \to t \tbar \bigr)
= 2 \pperp \sum_{ab} x_a x_b f_{a/A}(x_a) f_{b/B}(x_b)
{d\sigma (ab\to t\tbar)\over d\that}.}
The partonic cross section is folded together with distribution functions
$f_{a/A}(x_a)$ and $f_{b/B}(x_b)$ that specify the probability of finding
partons $a$ and $b$ inside hadrons $A$ and $B$ carrying momentum fractions
$x_a$ and $x_b$.  The product is then summed over initial parton
configurations.  The resulting hadronic cross section is a function of the
top and antitop rapidities $y_3$ and $y_4$ and their common transverse
momentum $\pperp$.

	The triply differential cross section in \ABttbarXsect\ may be
reduced to a function of a single variable by integrating over two of its
independent degrees of freedom.  We will concentrate upon the transverse
momentum differential cross section which we obtain by integrating
$d^3 \sigma / dy_3 dy_4 d\pperp$ over the rapidity range
$-2.5 \le y_3, y_4 \le 2.5$.
\foot{This rapidity range does not represent a fiducial cut but rather a
reasonable integration interval which contains the bulk of the produced
top quarks.  We have checked that extending the range to
$-6 \le y_3, y_4 \le 6$ does not noticibly alter our final results.}
The resulting $\pperp$ distribution of $t\tbar$ pairs produced at the LHC
is plotted in \LHCpTplot.
\foot{All of the results in \LHCpTplot\ and subsequent figures were calculated
using the next-to-leading order parton distribution function set B of Harriman,
Martin, Roberts and Stirling \Harriman\ evaluated at the renormalization scale
$\mu=\mperp\equiv\sqrt{\mtsq+\pperp^2}$.}
The solid curve in the figure illustrates the QCD differential cross section
$d\sigma_\QCD(PP \to t\tbar)/d\pperp$.  The dot-dashed, dashed and dotted
curves delineate the contributions from operators $\Osix_0$, $\Osix_1$ and
$\Osix_2$ that are generated after respectively setting $\Csix_0(\Lambda)=0.5$,
$\Csix_1(\Lambda)=0.5$ and $\Csix_2(\Lambda)=0.5$ in $\Leff$ with
$\Lambda=2\TeV$.  The QCD and nonrenormalizable operator curves must be added
together to obtain the effective field theory differential cross sections
$d\sigma_\EFT(PP \to t\tbar)/d\pperp$ that correspond to these nonzero
coefficient values.

	Several points about the results displayed in \LHCpTplot\ should be
noted.  Firstly, our choice for the operator coefficients is simply
representative.  Larger coefficient values lead to greater differences between
the QCD and EFT curves, while smaller values diminish the discrepancies.
Our particular choice for these parameters and the scale $\Lambda$ yields the
combined coefficient $\Csix_{0,1,2}(\Lambda)/\Lambda^2 = 0.5/(2 \TeV)^2
\simeq 4\pi/(10 \TeV)^2$ for the dimension-6 operators in effective Lagrangian
\effLagrange.  This value for the total coefficient is quite conservative
compared to that which has typically been used in previous quark substructure
studies.  Secondly, the transverse momentum
dependence of the curves in \LHCpTplot\ differentiates the dimension-6
operators from each other and the QCD terms in $\Leff$.  At low values of
$\pperp$, the contribution to $d\sigma(PP\to t\tbar)/d\pperp$ from the
magnetic moment operator $\Osix_0$ dominates over those from the triple
and double gluon field strength operators $\Osix_1$ and $\Osix_2$.  But the
$\Osix_0$ curve falls off much more rapidly with increasing transverse
momentum.  So by placing a lower $\pperp$ cut around 500 GeV, one can
eliminate most of the chromomagnetic moment operator's contribution while
retaining much of that from the double and triple field strength operators.
Finally, all the curves in the figure will be shifted around by higher
order QCD corrections.  The next-to-leading $O(\aS^3)$ corrections to the
tree level QCD differential cross section may be comparable to or even larger
than the $O(\aS^2)$ deviations induced by the dimension-6 terms in $\Leff$
depending upon their coefficients.  But the QCD and EFT distributions should
be compared at the same order in $\aS$.  Only then can deviations between them
be attributed to new physics beyond the Standard Model.

	It is interesting to compare the LHC differential cross sections in
\LHCpTplot\ with their Tevatron analogues shown in \FNALpTplot.  The Tevatron
curves were calculated at $\sqrt{s} = 1.8 \TeV$ using the same values for the
nonrenormalizable operator coefficients.  Not surprisingly, the total
integrated cross-section for $t\tbar$ production is two orders of magnitude
lower at the Tevatron than at the LHC.  Event rate is thus more of an issue
at the lower-energy machine.  We also clearly see from the two figures that
the relative importance of the dimension-6 terms in the effective
Lagrangian depends upon collider center-of-mass energy.  At the Tevatron,
$\Osix_2$ dominates over $\Osix_1$ for equal values of
their high energy scale coefficients.  This finding is intuitively reasonable
since the gluon content of colliding hadrons at $\sqrt{s}=1.8 \TeV$ is less
important than at $\sqrt{s}=14 \TeV$.

	The double gluon field strength operator affects quark scattering for
all flavors.  By comparing its predicted impact upon the inclusive jet cross
section with 1988 Tevatron data, we have previously set an upper bound
$|\Csix_2(\Lambda)|/\Lambda^2$ $\le 4\pi/(2 \TeV)^2$ on its coefficient
\ChoSimmons.  If $\Csix_2$ is allowed to assume this limiting value, we find
that $\Osix_2$ doubles the total integrated top quark cross section relative
to the lowest order QCD prediction.  This result is intriguing in light of the
recent CDF measurement
$\sigma(P\Pbar \to t\tbar)_{\rm CDF} / \sigma(P\Pbar \to t\tbar)_\QCD =
(13.9^{+6.1}_{-4.8} \pb)/(5.10^{+0.73}_{-0.43} \pb) \simeq 2.7^{+1.2}_{-1.0}$
\refs{\topmass,\Laenen}.  While it is premature to draw any conclusion
from this observation, we believe it is safe to say that the bound on
$\Csix_2$ should be significantly strengthened in the future by Tevatron
top quark data.

\newsec{Mapping the coefficient parameter space}

	In our effective field theory framework, the low energy effects of
any new strong interaction physics are encoded into the coefficients of the
nonrenormalizable terms in the effective Lagrangian.  These coefficients
define a multi-dimensional parameter space.  In order to assess the
likelihood of detecting signals of new strong sector physics at the LHC,
we need to map this space and determine the regions where deviations from
QCD could be measured.  Within those regions, we would then like to know
whether it is possible to isolate effects from individual operators in
$\Leff$.  We will explore these issues for the dimension-6 operators in our
basis in this section.

        To begin, we need to identify a measure of the difference between the
predictions of QCD and the strong interaction effective theory for top quark
pair production.  We will focus upon the disparities in their LHC transverse
momentum differential cross sections $d\sigma_\QCD(PP\to t\tbar)/d\pperp$ and
$d\sigma_\EFT(PP\to t\tbar)/d\pperp$.  It is important to recall that
experimental systematic errors will inevitably render uncertain the absolute
normalization for the observed $t\tbar$ distribution.  In future experimental
analyses, this normalization will undoubtedly be fitted to QCD at low
transverse momenta where any effects from new physics are expected to be
small.  We need to take this renormalization into account in our
theoretical analysis.  Therefore, we rescale by a multiplicative constant the
effective theory differential cross section which corresponds to the
distribution that will be experimentally measured.  We choose the constant
so that the renormalized $d\sigma_\EFT/d\pperp$ cross section coincides with
$d\sigma_\QCD/d\pperp$ at its maximum point.

        One simple choice for a dimensionless measure of the difference
between the QCD and EFT predictions for $d\sigma(PP \to t\tbar) /d\pperp$ is
the ratio of their integrals:
\eqn\ratio{\Rpperp = {\int d\pperp (d\sigma_\EFT/d\pperp) \over
\int d\pperp (d\sigma_\QCD/d\pperp)}.}
Since the disparity between $d\sigma_\QCD/d\pperp$ and $d\sigma_\EFT/d\pperp$
increases with $\pperp$, we perform the integrations in the numerator
and denominator of \ratio\ only over the high transverse momentum range
$500 \GeV \le \pperp \le 1000 \GeV$ in order to enhance the deviation of
$\Rpperp$ from unity.  The dependence of $\Rpperp$ upon the coefficient of
the chromomagnetic moment operator $\Osix_0$ is then much weaker than that for
gluonic operators $\Osix_1$ and $\Osix_2$.  So we plot $\Rpperp$ as a
function of $\Csix_1(\Lambda)$ and $\Csix_2(\Lambda)$ in \Rplot\ with all other
operator coefficients set equal to zero at the scale $\Lambda$.

	The origin in \Rplot\ necessarily lies along the $\Rpperp=1$ contour,
for the effective field theory reduces to QCD at this point.  It is clearly
offset from the center of the concentric contours displayed in the figure.
The offset is produced by the $O(1/\Lambda^2)$ interference terms in the
squared amplitude expressions \gluesqrdamp\ and \quarksqrdamp\ which are
linear in $\Csix_1$ and $\Csix_2$.  The smallness of the displacement in the
$\Csix_1$ direction demonstrates that the term proportional to
${\Csix_1}^2/\Lambda^4$ in $|A(gg \to t\tbar)|^2$ dominates over the
$\Csix_1/\Lambda^2$ interference term as we have previously discussed.

	While the ratio $\Rpperp$ provides a useful global measure of the
difference between the QCD and EFT transverse momentum $t\tbar$ distributions,
two points in the $\Csix_1$-$\Csix_2$ plane that lie along the same contour in
\Rplot\ may correspond to two very different $d\sigma(PP\to t\tbar)/d\pperp$
curves.  It is therefore instructive to
consider a second dimensionless measure which is sensitive to the curves'
shapes.  To construct such a quantity, we first discretize the
transverse momentum interval $500 \GeV \le \pperp \le 1000 \GeV$ into $N=20$
bins.  We then multiply the value for $d\sigma/d\pperp$ in each bin by the
binwidth $\Delta\pperp$, the integrated luminosity $\int\CL dt$, the branching
ratio BR for the $t\tbar$ pair's single lepton plus jets decay mode, and the
$b$-tagging efficiency $\e_b$ to convert the differential cross section
into a corresponding number of observable $t\tbar$ events:
\eqn\Ni{N_i = \Bigl({d\sigma \over d\pperp} \Bigr)_i \times \Delta \pperp
\times {\int \CL dt} \times {\rm BR} \times \e_b.}
We shall take the numerical values for these parameters to be $\Delta\pperp
= 25 \GeV$, $\int\CL dt = 30 \> {\rm fb}^{-1}$, ${\rm BR} = 24/81$ and
$\e_b=0.25$.  Finally, we quantify the difference between the effective
theory distribution which will be measured for nonvanishing coefficient
values and the theoretical QCD prediction in terms of a $\chisq$
function.  The highest $\pperp$ bins are the most important for
discriminating between the two distributions, but they contain the
fewest events.  So we adopt the Poisson $\chisq$ function
\eqn\Poissonchisq{\chisq = 2 \sum_{i=1}^N \Bigl[ N_i^\QCD - N_i^\obs + N_i^\obs
 \ln {N_i^\obs \over N_i^\QCD}\Bigr]}
which is appropriate for low statistics \PDB.

	We plot $\chisq/N$ in \chisqplot\ as a function of $\Csix_1(\Lambda)$
and $\Csix_2(\Lambda)$ over the same region of coefficient parameter space
as in \Rplot.  The innermost crescent contour in the figure corresponds to
the expectation value $\chisq/N=1$.  For points lying within this contour, the
probability that an observed transverse momentum differential cross section
could be attributed to a statistical fluctuation of QCD rather than to
nonvanishing values for the $\Csix_1(\Lambda)$ and $\Csix_2(\Lambda)$
coefficients is greater than 50 \%.  As QCD and the strong interaction
effective theory cannot be meaningfully told apart inside the first contour,
its boundary establishes a limit on the sensitivity to new gluon sector
physics which can be achieved at the
LHC.  The surrounding contours in \chisqplot\ illustrate selected $\chisq/N$
standard deviation levels where $\sigma=\sqrt{2/N}=0.316$ for $N=20$ degrees
of freedom.  For example, the gluonic operator coefficients lying on the
outermost contour yield $d\sigma(PP \to t\tbar)_\EFT/d\pperp$ distributions
which can be distinguished from $d\sigma(PP\to t\tbar)_\QCD/d\pperp$ at the
$8 \sigma$ level.  If the horizontal and vertical axes were drawn to the same
scale in \chisqplot, this last contour would appear as an ellipse
approximately three times more narrow in the $\Csix_1(\Lambda)$ direction than
in the $\Csix_2(\Lambda)$ direction.  The $\chisq/N$ curves thus quantify the
extent to which future LHC experiments will be more sensitive to the gluonic
$G^3$ operator than to its four-quark competitors.

	The information contained within the contour plots of \Rplot\ and
\chisqplot\ is insufficient to completely determine where an observed
$d\sigma(PP \to t\tbar)/d\pperp$ function lies within the $\Csix_1$-$\Csix_2$
parameter space.  But taken together, the two graphs significantly restrict
the allowed values for these coefficients.  Clearly, other differences between
QCD and the effective field theory can be investigated along the lines which
we have followed here.  In particular, their predictions for the angular
distributions of $t\tbar$ pairs can further constrain the allowed region
within the coefficient parameter space.  We briefly touch on this topic in
the next section.

\newsec{Top-antitop angular distributions}

	Our study of the prospects for detecting new strong interaction
physics at the LHC has so far utilized only the transverse momentum
information incorporated within $d^3 \sigma(PP \to t\tbar)/dy_3 dy_4
d\pperp$.  The triply differential cross section contains, however,
complementary angular distribution information.  We consider its implications
for discriminating between the nonrenormalizable operators within the
effective Lagrangian in this section.

	A number of angular distributions for $t\tbar$ pairs as well as
their decay products can be generated by integrating $d^3 \sigma/dy_3
dy_4 d\pperp$ over various ranges of $y_3$, $y_4$ and $\pperp$.  One
differential cross section of particular interest is
$d\sigma(PP \to t\tbar)/d \costhstar$ where $\theta^*$ denotes the angle
between the direction of the boost and that of the top quark in the parton
center-of-mass frame. We plot this cross section in \angleplot\ for pure QCD
and QCD plus some of the nonrenormalizable operators in our basis set.
In order to enhance the
operators' signal over the QCD background, we have imposed the transverse
momentum cut $\pperp \ge 500 \GeV$.  We have also required the
lab frame angle between the $t$ or $\tbar$ and the beamline to exceed
$25.4^\circ$ to approximate the acceptance of planned LHC detectors.  This
last restriction ensures that the pseudorapidities of the decay products from
high momentum tops will predominantly lie within the interval
$-2.5 \le \eta \le 2.5$.

	The solid curve in \angleplot\ represents the QCD differential
cross section $d\sigma(PP \to t\tbar)_\QCD/d\costhstar$.  The dotted,
dashed and dot-dashed curves in the figure illustrate
$d\sigma(PP \to t\tbar)_\EFT/d\costhstar$ for $\Csix_1(\Lambda) = 0.5$,
$\Ceight_3(\Lambda) = 0.5$, and $\Ceight_3(\Lambda) = - 0.5$ with
$\Lambda = 2 \TeV$.
\foot{The coefficient $\Ceight_3$ was not evolved using the renormalization
group but was instead simply fixed at its $\Lambda$ scale value.}
We again find that the effect of the triple gluon field
strength operator $\Osix_1$ is essentially independent of the sign of its
coefficient.  The differential cross section corresponding to
$\Csix_1(\Lambda)=-0.5$ thus closely traces that for $\Csix(\Lambda)=0.5$
displayed in the figure.  The dimension-8 gluon operator $\Oeight_3$ induces
deviations from pure QCD which are clearly visible in $d\sigma/d\costhstar$.
This result is both surprising and interesting since we previously found that
the effect of $\Oeight_3$ upon the $t\tbar$ transverse momentum distribution
was negligible.  Indeed, we did not include a curve corresponding to the
dimension-8 gluon operator in the transverse momentum plots of \LHCpTplot\
and \FNALpTplot\ since it would have been suppressed relative to its
dimension-6 counterpart by more than an order of magnitude.  But
we now see that experiments at the LHC can be sensitive to this
next-to-next-to-leading order operator if they probe its impact on the
$t\tbar$ angular distribution.

	As for the transverse momentum distributions, it is again useful to
identify dimensionless measures of the differences between the QCD and EFT
predictions for $d\sigma(PP \to t\tbar)/d\costhstar$.  One simple choice for
such a measure is the ratio of their integrals
\eqn\Rangdefn{\Rang = {\int d\costhstar (d\sigma_\EFT/d\costhstar) \over
\int d\costhstar (d\sigma_\QCD/d\costhstar)}}
which is the analog of $\Rpperp$ in eqn.~\ratio.  Another is
the ratio $\Rrms$ of their root-mean-squared values for $\costhstar$.
We have calculated these ratios for the angular differential cross sections
shown in \angleplot\ and for similar cross sections involving the magnetic
moment and four-quark operators.  For the curves corresponding to
$\Csix_0(\Lambda)$, $\Csix_1(\Lambda)$, $\Csix_2(\Lambda)$ and
$\Ceight_3(\Lambda)$ equal to 0.5, we find $\Rang = (1.03, 1.23, 1.46, 0.82)$
and $\Rrms = (0.999, 0.978, 0.991, 0.871)$.  For analogous curves
with the $\Lambda$ scale coefficients set equal to -0.5, we find
$\Rang = (0.969, 1.23, 0.735, 1.18)$ and $\Rrms = (1.00, 0.978, 1.03, 1.08)$.
The most striking conclusion which we draw from these results is that the
dimension-8 gluon operator alters the shape of the $t\tbar$ angular
distribution much more than all the other dimension-6 operators in $\Leff$ for
comparable values of their coefficients.  The magnetic moment
operator's angular distribution is indistinguishable from that of pure
QCD, while the distributions of the $G^3$ and $(DG)^2$ operators differ
significantly from that of QCD in $\Rang$ but not in $\Rrms$.
$\Oeight_3$ thus possesses a distinctive signature: a QCD-like
transverse momentum distribution but a quite nonstandard angular
distribution with $\Rang$ and $\Rrms$ both deviating from unity in the
same direction.

	A more precise determination of the power of angular measurements to
delimit allowed regions of coefficient parameter space will require
detailed simulations including top quark decays and detector resolution.
We leave such a study to future work.

\newsec{Conclusions}

	In this article, we have investigated the impact of the triple gluon
field strength operator upon top quark pair production.  The $G^3$ operator
represents the only genuinely gluonic CP even term which can arise at
dimension-6 within an effective strong interaction Lagrangian.  Although it
has proven surprisingly difficult to study in the past, the prospects for
either detecting or significantly constraining this operator appear quite
promising at the LHC where its effects will be significantly enhanced by the
large gluon content at small $x$ of the colliding hadrons.  We have found that
the sensitivity of the $t\tbar$ transverse momentum distribution to the triple
field strength operator is greater than that to all other dimension-6 pure
quark and mixed quark-gluon terms in $\Leff$ for comparable values of their
coefficients.  Moreover, only one higher order gluonic operator can contribute
to $gg \to t\tbar$ at dimension-8, and its effect on the $\pperp$ distribution
is more than an order of magnitude smaller than that of $G^3$.  We have also
seen that angular distribution information can help to differentiate
effects from the various operators which may reside within the effective
Lagrangian.  Top-antitop production therefore promises to provide an important
means for probing new strong interaction physics beyond the Standard Model.

\bigskip\bigskip
\centerline{\bf Acknowledgments}
\bigskip

      It is a pleasure to thank Tao Han, Kenneth Lane, Rolf Mertig, Stephen
Mrenna and Bing Zhou for helpful discussions.

\listrefs
\listfigs
\bye